\documentclass[10pt,a4paper,notitlepage]{article}
\usepackage[latin1]{inputenc}
\usepackage{graphicx,amsmath,amsfonts,amssymb,mathrsfs}

\def\arraystretch{1.5}

\title{Dynamical Symmetries in\\Supersymmetric Matrix Models\footnote{Supported by the Swedish Research Council}}
\author{
	\renewcommand{\thefootnote}{\arabic{footnote}}
	V. Bach\footnotemark[1],\quad 
	J. Hoppe\footnotemark[2],\quad 
	D. Lundholm\footnotemark[3]
}
\date{}

\begin{document}

\maketitle
\renewcommand{\thefootnote}{\arabic{footnote}}
\footnotetext[1]{vbach@mathematik.uni-mainz.de}
\footnotetext[2]{hoppe@math.kth.se}
\footnotetext[3]{dogge@math.kth.se}
\renewcommand{\thefootnote}{\arabic{footnote}}

\begin{abstract}
	We reveal a dynamical SU(2) symmetry in the asymptotic description of 
	supersymmetric matrix models. We also consider a recursive approach
	for determining the ground state, and point out some 
	additional properties of the model(s).
\end{abstract}

\section{Introduction}

	Ten years ago, a lot of effort was put into the question of zero-energy 
	states in SU($N$)-invariant supersymmetric matrix models. While the attempt
	to explicitly construct such a state mainly \cite{hoppe} used a 
	space-independent (but manifest SO($d$)-invariance-breaking)
	decomposition of the fermions into creation and 
	annihilation operators, 
	the asymptotic form of the wave function was
	determined with the help of space-dependent fermions 
	\cite{halpern_schwartz, graf_hoppe, froehlich_et_al}.

	In this paper we would like to point out a dynamical $\textrm{SU}(2)$ symmetry 
	and the formation of `Cooper pairs' that arise in the $\textrm{SO}(d)$-breaking 
	formulation when diagonalizing certain ingredients of the fermionic
	part of the Hamiltonian.
	We start by considering the asymptotic $\textrm{SU}(2)$ theory
	but note that several features extend to the non-asymptotic and general 
	$\textrm{SU}(N)$ cases.

\section{Asymptotic form of the Hamiltonian}

	The bosonic configuration space is a set of $d$ = 2, 3, 5, or 9 traceless hermitian
	matrices $X_s$, corresponding to the Lie algebra of the gauge group SU($N$).
	For simplicity, we start by taking $N=2$.
	Coordinatizing regions where the bosonic potential,
	\begin{equation} \label{potential}
		V = - \frac{1}{2} \sum_{s,t=1}^{d} \textrm{tr}\ [X_s,X_t]^2,
	\end{equation}
	is zero by (cp. e.g. \cite{graf_hoppe, froehlich_et_al})
	\begin{equation} \label{coords}
	\begin{array}{rcl}
		X_t &=& r \cos\theta\ \tilde{E}_t\ (\frac{1}{2} e_A \sigma_A), \quad t=1,\ldots,d-2 \\
		X_{d-1} + iX_d &=& r \sin\theta\ e^{i\varphi}\ (\frac{1}{2} e_A \sigma_A),
	\end{array}
	\end{equation}
	where $\boldsymbol{e}^2 = 1, \sum_{t=1}^{d-2} \tilde{E}_t^2 = 1$ and $\sigma_A$ 
	are the Pauli matrices, $\frac{1}{r}$ times
	the effective asymptotic Hamiltonian (cf. \cite{froehlich_et_al, hoppe}) 
	becomes	
	\begin{equation} \label{hamiltonian}
	\begin{array}{rcl}
		H^\infty = H^\infty_B 
			&+& 2\cos \theta\ (-ie_C\epsilon_{ABC})\ \Gamma_{\alpha\beta}\ \lambda_{\alpha A} \partial_{\lambda_{\beta B}} \\
			&+& \sin \theta\ e^{i\varphi}\ (e_C\epsilon_{ABC})\ \lambda_{\alpha A} \lambda_{\alpha B} \\
			&+& \sin \theta\ e^{-i\varphi}\ (e_C\epsilon_{ABC})\ \partial_{\lambda_{\alpha B}} \partial_{\lambda_{\alpha A}}.
	\end{array}
	\end{equation}
	The last three terms are the leading ones in the fermionic part of
	the Hamiltonian (as $r \to \infty$) while
	$H_B^\infty$,
	which arises from $-\Delta + V$ in that limit,
	denotes an independent set of harmonic oscillators on $\mathbb{R}^{s_d}$
	with ground state energy $s_d$ = 2, 4, 8, or 16.
	$\Gamma := \sum_{t=1}^{d-2} \tilde{E}_t \Gamma_t$ is a
	purely imaginary, antisymmetric, and hence self-adjoint
	$\frac{s_d}{2} \times \frac{s_d}{2}$-matrix
	squaring to unity.
	$\lambda_{\alpha A}$ and $\partial_{\lambda_{\alpha A}} = \lambda^\dagger_{\alpha A}$ 
	are space-independent fermion creation resp. annihilation operators satisfying
	\begin{equation} \label{indep_anticomm_rels}
	\begin{array}{c}
		\{ \lambda_{\alpha A}, \partial_{\lambda_{\beta B}} \} = \delta_{\alpha \beta} \delta_{AB}, \\
		\{ \lambda_{\alpha A}, \lambda_{\beta B} \} = \{ \partial_{\lambda_{\alpha A}}, \partial_{\lambda_{\beta B}} \} = 0
	\end{array}
	\end{equation}
	and acting on the fermionic vacuum state $|0\rangle$, 
	defined by $\partial_{\lambda_{\alpha A}} |0\rangle = 0 \ \forall \alpha, A$.

	We define space-dependent fermion creation operators (for $d$ = 3, 5, 9)
	\begin{equation} \label{dep_lambdas}
		\lambda^{\phantom{\dagger}}_{\sigma j \tau} := (\tilde{e}_{\sigma j})^{\phantom{\dagger}}_\alpha (\boldsymbol{n}_\tau)^{\phantom{\dagger}}_A \lambda^{\phantom{\dagger}}_{\alpha A},
	\end{equation}
	where $\sigma, \tau$ denote + or -, and $\boldsymbol{n}_\pm \in \mathbb{C}^3$ 
	resp. $\tilde{e}_{\sigma j} \in \mathbb{C}^{s_d/2}$ are eigenvectors of
	$(-i e_C \epsilon_{CAB})$ resp. $\Gamma_{\alpha\beta}$,
	\begin{equation} \label{eigenvectors_n}
		i \boldsymbol{e} \times \boldsymbol{n}_\pm = \pm \boldsymbol{n}_\pm
	\end{equation}
	\begin{equation} \label{eigenvectors_e}
		\Gamma \tilde{e}_{\pm j} = \pm \tilde{e}_{\pm j}, \quad j = 1,\ldots,\frac{s_d}{4}.
	\end{equation}
	We choose these to depend continuously on $\boldsymbol{e}$ and $\tilde{E}$, as well as to be
	orthonormal and such that the complex conjugates
	$(\boldsymbol{n}_\pm)^* = \boldsymbol{n}_\mp$ 
	and $(\tilde{e}_{\pm j})^* = \tilde{e}_{\mp j}$.
	The asymptotic Hamiltonian $H^\infty$, when acting on the ground state 
	of $H^\infty_B$, can then be written as
	\begin{equation} \label{hamiltonian_split}
		\textstyle
		H = H_0 + H_+ + H_-,
	\end{equation}
	\begin{equation} \label{hamiltonian_parts}
	\begin{array}{lcl}
		H_0 &=& s_d + 2\cos \theta\ \sum_j (N_{A_j} - N_{B_j}), \\
		H_+ &=& 2\sin \theta\ \sum_j (A^{\phantom{\dagger}}_j + B^{\phantom{\dagger}}_j), \\
		H_- &=& 2\sin \theta\ \sum_j (A^{\dagger}_j + B^{\dagger}_j), \\
	\end{array}
	\end{equation}
	where
	\begin{equation} \label{cooper}
	\begin{array}{rcl}
		A^{\phantom{\dagger}}_j &:=& ie^{i\varphi} \lambda^{\phantom{\dagger}}_{+j+} \lambda^{\phantom{\dagger}}_{-j-}, \\
		B^{\phantom{\dagger}}_j &:=& ie^{i\varphi} \lambda^{\phantom{\dagger}}_{-j+} \lambda^{\phantom{\dagger}}_{+j-},
	\end{array}
	\end{equation}
	satisfy
	\begin{equation} \label{cooper_number}
	\begin{array}{c}
		\ [ A^{\phantom{\dagger}}_j, A^{\dagger}_j ] = N_{A_j}-1 := \lambda_{+j+} \partial_{\lambda_{+j+}} + \lambda_{-j-} \partial_{\lambda_{-j-}} - 1, \\ 
		\ [ B^{\phantom{\dagger}}_j, B^{\dagger}_j ] = N_{B_j}-1 := \lambda_{-j+} \partial_{\lambda_{-j+}} + \lambda_{+j-} \partial_{\lambda_{+j-}} - 1.
	\end{array}
	\end{equation}
	For the $d$=2 case we instead of \eqref{dep_lambdas} define 
	$\lambda_\pm := (\boldsymbol{n}_\pm)^{\phantom{\dagger}}_A \lambda^{\phantom{\dagger}}_{A}$
	and the corresponding expressions for the asymptotic Hamiltonian are simply
	\begin{equation} \label{hamiltonian_2d}
		H_0 = 2, \quad 
		H_+ = 2 C, \quad 
		H_- = 2 C^\dagger, \quad 
		C := ie^{i\varphi} \lambda^{\phantom{\dagger}}_{+} \lambda^{\phantom{\dagger}}_{-}.
	\end{equation}

\section{Dynamical symmetry}

	Let us now restrict to $d$=9 (for definiteness).
	Denoting $A := \sum_j A_j$ by
	$J_+ \otimes 1$, $A^\dagger = \sum_j A^\dagger_j$ by $J_- \otimes 1$, 
	$\frac{1}{2}(N_{A} - 4) := \frac{1}{2}(\sum_j N_{A_j} - 4)$ 
	by $J_3 \otimes 1$, 
	and similarly $1 \otimes J_+, 1 \otimes J_-, 1 \otimes J_3$ for the $B$s, with
	\begin{equation} \label{j_relations}
		[J_+,J_-] = 2J_3, \quad
		[J_3,J_\pm] = \pm J_\pm, \quad
		J_\pm = J_1 \pm iJ_2,
	\end{equation}
	eqs. \eqref{hamiltonian_split}, \eqref{hamiltonian_parts} can be written as
	\begin{equation} \label{hamiltonian_j}
		\textstyle
		\frac{1}{4} H = (2 + \cos\theta J_3 + \sin\theta J_1) \otimes 1
			\ +\  1 \otimes (2 - \cos\theta J_3 + \sin\theta J_1),
	\end{equation}
	thus exhibiting the dynamical symmetry mentioned above.
	The relevant SU(2) representations are the tensor product of four spin $\frac{1}{2}$ 
	representations, i.e. direct sums of two singlets 
	(note that both $(A_1A_3+A_2A_4-A_1A_4-A_2A_3)|0\rangle$ and $(A_1A_2+A_3A_4-A_1A_4-A_2A_3)|0\rangle$ 
	are annihilated by $A$, $A^\dagger$, and $\frac{1}{2}(N_A-4)$), 
	three spin 1 representations, and
	(most importantly, as providing the zero-energy state of $H$) one spin 2 representation
	acting irreducibly on the space spanned by the orthonormal states
	\begin{equation} \label{spin_2_space}
		\textstyle
		|0\rangle, \quad 
		\frac{1}{2}A|0\rangle, \quad
		\frac{1}{\sqrt{24}}A^2|0\rangle, \quad
		\frac{1}{12}A^3|0\rangle, \quad
		\frac{1}{4!}A^4|0\rangle.
	\end{equation}
	Restricting to that space (correspondingly for the $B$s), we can write
	\begin{equation} \label{j_matrices}
		\def\arraystretch{1.0}
		J_3 = \left[
		\begin{array}{ccccc}
			\!-2& 0& 0& 0& 0 \\
			 0&\!-1& 0& 0& 0 \\
			 0& 0& 0& 0& 0 \\
			 0& 0& 0& 1& 0 \\
			 0& 0& 0& 0& 2
		\end{array}
		\right], \quad
		J_+ = \left[
		\begin{array}{ccccc}
			 0& 0& 0& 0& 0 \\
			 2& 0& 0& 0& 0 \\
			 0& \sqrt{6}& 0& 0& 0 \\
			 0& 0&\!\sqrt{6}& 0& 0 \\
			 0& 0& 0& 2& 0
		\end{array}
		\right], \quad
		J_- = \left[
		\begin{array}{ccccc}
			 0& 2& 0& 0& 0 \\
			 0& 0& \sqrt{6}& 0& 0 \\
			 0& 0& 0&\!\sqrt{6}& 0 \\
			 0& 0& 0& 0& 2 \\
			 0& 0& 0& 0& 0
		\end{array}
		\right].
	\end{equation}
	Since the spectrum of $\sin\theta J_1 \pm \cos\theta J_3$ is the same
	as that of $J_3$, the spectrum of $\frac{1}{4}H$ 
	clearly consists of all integers between zero and eight, with the unique zero-energy
	state $\Psi$ most easily obtained by solving individually,
	for each $A_j$ resp. $B_j$ degree of freedom,
	\begin{equation} \label{ind_groundstate_eq}
		\left(1 \pm \cos\theta\ \sigma_3^{(j)} + \sin\theta\ \sigma_1^{(j)}\right) \Psi
		= e^{\mp \frac{1}{2} \theta i\sigma_2^{(j)}} \left(1 \pm \sigma_3^{(j)}\right) e^{\pm \frac{1}{2} \theta i\sigma_2^{(j)}} \Psi
		\stackrel{!}{=} 0,
	\end{equation}
	where we identify 
	$2J_k = \sigma_k^{(1)} \otimes 1 \otimes 1 \otimes 1 + \ldots + 1 \otimes 1 \otimes 1 \otimes \sigma_k^{(4)} = \sum_{j=1}^4 \sigma_k^{(j)}$.
	In our notation 
	$\sigma_3^{(j)} |0\rangle = -|0\rangle$ and $\sigma_3^{(j)} A_j|0\rangle = +A_j|0\rangle$, 
	and we easily find the solution to \eqref{ind_groundstate_eq} as
	\begin{equation} \label{groundstate_rot}
		\textstyle
		\Psi
			= \left( \prod_j e^{-\frac{\theta}{2}i\sigma_2^{(j)}} \right) \left( \prod_j e^{\frac{\theta}{2}i\sigma_2^{(j)}} B_j \right) |0\rangle \\
			= e^{-\theta i(J_2 \otimes 1 - 1 \otimes J_2)} \frac{B^4}{4!} |0\rangle.
	\end{equation}
	Using the nilpotency of $A_j$ and $B_j$ for
	\begin{equation} \label{exponential_relations}
		e^{ \alpha(A_j-A^\dagger_j)} |0\rangle = \cos\alpha\ e^{\tan\alpha \phantom{.} A_j} |0\rangle 
		\quad \textrm{and} \quad
		e^{-\alpha(B_j-B^\dagger_j)} B_j |0\rangle = \sin\alpha\ e^{\cot\alpha \phantom{.} B_j} |0\rangle,
	\end{equation}
	the ground state can also be written as
	\begin{equation} \label{groundstate_stat}
	\begin{array}{rcl}
		\Psi 
			&=& \frac{1}{16} e^{-4i\varphi} (\sin\theta)^{-4} \prod_j \big( \sin\theta - (1-\cos\theta)A_j \big) \big( \sin\theta - (1+\cos\theta)B_j \big) |0\rangle \\
			&=& \frac{1}{16} e^{-4i\varphi} (\sin\theta)^4 e^{-\frac{1-\cos\theta}{\sin\theta} A - \frac{1+\cos\theta}{\sin\theta} B} |0\rangle \sim e^{-C_\theta} |0\rangle,
	\end{array}
	\end{equation}
	with $C_\theta := \frac{1-\cos\theta}{\sin\theta} (J_+ \otimes 1) + \frac{1+\cos\theta}{\sin\theta} (1 \otimes J_+)$.
	Alternatively, one can solve the $2 \times 2$ matrix eigenvector 
	equations resulting from \eqref{ind_groundstate_eq},
	\begin{equation} \label{ind_groundstate_eqs}
	\begin{array}{c}
		\left(1 + \cos\theta (N_{A_j}-1) + \sin\theta (A_j + A^\dagger_j)\right) \Psi = 0, \\
		\left(1 - \cos\theta (N_{B_j}-1) + \sin\theta (B_j + B^\dagger_j)\right) \Psi = 0,
	\end{array}
	\end{equation}
	to obtain \eqref{groundstate_stat}.

	For $d$=2 the asymptotic ground state is easily found from \eqref{hamiltonian_2d},
	\begin{equation} \label{groundstate_2d}
		\textstyle
		\Psi
			= \frac{1}{\sqrt{2}} e^{-C} |0\rangle
			= \frac{1}{\sqrt{2}} (1-C) |0\rangle.
	\end{equation}
	
	An interesting feature of the form \eqref{groundstate_rot} for the 
	ground state is that it expresses it as a spin-rotation by an angle
	$\theta$ applied to some reference state $B^4|0\rangle$
	(which itself also varies in the first $d-2$ directions in space
	according to \eqref{coords}, \eqref{dep_lambdas}, \eqref{eigenvectors_e}).

\section{Graded chain of Hamiltonians}

	Consider the grade- resp. fermion number-ordered equations
	\begin{equation} \label{graded_eqs}
	\begin{array}{lclcll}
						 && H_0 \Psi_0 		&+& H_- \Psi_2 		&= 0, \\
		H_+ \Psi_0 		&+& H_0 \Psi_2 		&+& H_- \Psi_4 		&= 0, \\
		H_+ \Psi_2 		&+& H_0 \Psi_4		&+& H_- \Psi_8 		&= 0, \\
		&&\vdots \\
		H_+ \Psi_{12}	&+& H_0 \Psi_{14}	&+& H_- \Psi_{16}	&= 0, \\
		H_+ \Psi_{14}	&+& H_0 \Psi_{16}	 && 				&= 0, \\
	\end{array}
	\end{equation}
	implied by $H\Psi = (H_0 + H_+ + H_-)(\Psi_0 + \Psi_2 + \ldots + \Psi_{16}) \stackrel{!}{=} 0$.
	(We have dropped the eight non-dynamical parallel fermions 
	$\lambda^{\parallel}_\alpha := \lambda_{\alpha A} e_A$.)
	The following method to construct the ground state we believe
	to be relevant also for the fully interacting, non-asymptotic theory.
	Use the first equation in \eqref{graded_eqs} to express $\Psi_0$ in terms
	of $\Psi_2$,
	\begin{equation} \label{recursion_first}
		\Psi_0 = -H_0^{-1} H_- \Psi_2.
	\end{equation}
	$H_0$ is certainly invertible on the zero-fermion subspace, 
	even in the full theory, where (cf. \cite{hoppe})
	\begin{equation} \label{full_h0}
		H_0 = -\Delta + V - 2i x_{jC} f_{CAB} \Gamma_{\alpha\beta}^j  \lambda^{\phantom{\dagger}}_{\alpha A} \lambda^{\dagger}_{\beta B}.
	\end{equation}
	Using \eqref{recursion_first}, the second equation in \eqref{graded_eqs}
	can be written as
	\begin{equation} \label{grade_two_eq}
		H_2 \Psi_2 + H_- \Psi_4 = 0, \quad \textrm{with} \quad H_2 := H_0 - H_+ H_0^{-1} H_-,
	\end{equation}
	yielding
	\begin{equation} \label{recursion_second}
		\Psi_2 = -H_2^{-1} H_- \Psi_4,
	\end{equation}
	provided $H_2$ is invertible on $H_-\Psi_4$, resp. the two-fermion sector
	of the Hilbert space.
	Continuing in this manner, denoting
	\begin{equation} \label{subspace_2k}
		\hat{\mathscr{H}}_{2k} := \textrm{Span} \big\{ A^{m} B^{n} |0\rangle \big\}_{m,n=0,1,2,3,4,\ m+n=k}
	\end{equation}
	for the considered $2k$-fermion subspace, 
	we find that if we assume invertibility of $H_{2k}$ on $\hat{\mathscr{H}}_{2k}$ we can form
	\begin{equation} \label{hamiltonian_2k}
		H_{2(k+1)} := H_0 - H_+ H_{2k}^{-1} H_- 
	\end{equation}
	on $\hat{\mathscr{H}}_{2(k+1)}$ and solve for $\Psi_{2k}$ 
	in terms of $\Psi_{2(k+1)}$. 
	The final equation for $\Psi_{16}$ is $H_{16} \Psi_{16} = 0$.
	
	For concreteness, denote an orthonormal basis of $\hat{\mathscr{H}} = \oplus_k \hat{\mathscr{H}}_{2k}$
	by $|k,l\rangle := |k\rangle \otimes |l\rangle$, where,
	as in \eqref{spin_2_space},
	\begin{equation} \label{spin_2_basis_state}
		|k\rangle := \frac{1}{ k! \sqrt{\binom{4}{k}} } J_+^k |0\rangle.
	\end{equation}
	Then $H_+H_0^{-1}H_-$, e.g., acts on $\hat{\mathscr{H}}$ `tridiagonally' according to
	\begin{equation} \label{tridiagonal}
	\def\arraystretch{2.0}
	\begin{array}{ll}
		\frac{1}{\sin^2\theta} H_+H_0^{-1}H_- |k,l\rangle 
			&= \left( \frac{k(5-k)}{4 + (k-l-1)\cos\theta} + \frac{l(5-l)}{4 + (k-l+1)\cos\theta} \right) |k,l\rangle \\
			&\quad +\ \frac{\sqrt{l(5-l)(k+1)(4-k)}}{4 + (k-l+1)\cos\theta} |k+1,l-1\rangle \\
			&\quad +\ \frac{\sqrt{k(5-k)(l+1)(4-l)}}{4 + (k-l-1)\cos\theta} |k-1,l+1\rangle.
	\end{array}
	\end{equation}
	
	Calculating the spectra of $H_{2k}$ on $\hat{\mathscr{H}}_{2k}$ (e.g. with the help of a computer)
	one can verify the invertibility of all $H_{2k}$ on $\hat{\mathscr{H}}_{2k}$ for $k<8$,
	while $H_{16}$ is identically zero on $\hat{\mathscr{H}}_{16}$.
	Hence, one can also start with the state $\Psi_{16} \sim A^4B^4|0\rangle$
	(with correct normalization in $\theta$) and generate the 
	lower grade parts of the full ground state $\Psi$ using the relations
	\eqref{recursion_first}, \eqref{recursion_second}, etc.

	Let us finish this section by noting a simple consequence of 
	the graded form \eqref{graded_eqs}
	of the ground state equation $H\Psi=0$ (for general $d$ and $N$). 
	Taking the inner product of the grade $2k$-equation with $\Psi_{2k}$
	yields 
	\begin{equation} \label{exp_value_recursion}
	\begin{array}{ll}
		\langle \Psi_{2k}, H_-\Psi_{2(k+1)} \rangle 
		&= -\langle H_0 \rangle_{2k} - \langle \Psi_{2k}, H_+\Psi_{2(k-1)} \rangle \\
		&= -\langle H_0 \rangle_{2k} - \langle \Psi_{2(k-1)}, H_-\Psi_{2k} \rangle^*,
	\end{array}
	\end{equation}
	where $\langle H_0 \rangle_{2k} := \langle \Psi_{2k}, H_0\Psi_{2k} \rangle$.
	The first equation reads $\langle \Psi_0, H_-\Psi_2 \rangle = -\langle H_0 \rangle_0$
	which is real.
	The second then becomes $\langle \Psi_2, H_-\Psi_4 \rangle = -\langle H_0 \rangle_2 + \langle H_0 \rangle_0$,
	and so on, so that in the last step one obtains
	\begin{equation} \label{exp_value_relation}
		\sum_{k=0}^{\Lambda} (-1)^k \langle H_0 \rangle_{2k} = 0,
	\end{equation}
	where $\Lambda$ is the total number of fermions in the relevant Fock space.

	It is instructive to verify \eqref{exp_value_relation} for the
	asymptotic $N=2$ case studied above, since there all relevant terms can be
	calculated explicitly. Using the basis \eqref{spin_2_basis_state}
	and the notation $\alpha := 1-\cos\theta$, $\beta := 1+\cos\theta$, we find
	\begin{equation} \label{groundstate_in_spin_2_basis}
		\Psi \sim e^{-C_\theta} |0\rangle = 
		\sum_{k} \frac{ (-1)^k \sqrt{\binom{4}{k}} }{(\sin\theta)^k} \alpha^k |k\rangle \otimes
		\sum_{l} \frac{ (-1)^l \sqrt{\binom{4}{l}} }{(\sin\theta)^l} \beta^l |l\rangle.
	\end{equation}
	Hence,
	\begin{equation} \label{groundstate_exp_value}
		\langle \Psi_{2n}, H_0\Psi_{2n} \rangle = \frac{1}{64} (\sin\theta)^{8-2n} \sum_{k+l=n} \binom{4}{k} \binom{4}{l} \left(4 + (k-l)\cos\theta\right) \alpha^{2k} \beta^{2l}.
	\end{equation}

\section{General SU($N$)}

	Let us now derive, for general $N \geq 2$, the ground state energy of 
	\begin{equation} \label{h_f}
		H_F = i \gamma^t_{\alpha\beta} f_{ABC} x_{tC} \theta_{\alpha A} \theta_{\beta B}
	\end{equation}
	in regions of the configuration space where the potential $V$ is zero.
	(As in \eqref{full_h0}, $f_{ABC}$ denote the structure constants of 
	SU($N$) in an orthonormal basis.)
	By \eqref{potential} this means that all $X_s$ are commuting, hence can be written
	$X_s = U D_s U^\dagger$ where $U$ is unitary and independent of $s$ and the $D_s$
	are diagonal. If we look into a particular direction (corresponds to fixing 
	$\boldsymbol{e}$ in the SU(2) case) and choose a basis $\{T_A\}$ accordingly we may write
	$X_s = D_s = x_{sA} T_A = x_{s\tilde{k}} T_{\tilde{k}}$ and $x_{sa} = 0$, where 
	$\tilde{k} = 1,\ldots,N-1$ are indices in the Cartan subalgebra and $a,b = N,\ldots,N^2-1$
	denote the remaining indices.
	
	Denoting the eigenvalues of $X_t$ by $\mu^t_k$, i.e. $X_t = \textrm{diag}(\mu^t_1,\ldots,\mu^t_N)$,
	then the eigenvectors $\{ e_{kl} \}_{k \neq l}$ of
	$M^t_{ab} := -i f_{abC} x_{tC} = -i f_{ab\tilde{k}} x_{t\tilde{k}}$
	satisfy (cf. e.g. \cite{hasler_hoppe})
	\begin{equation} \label{mt_eigenvectors}
		M^t e_{kl} = (\mu^t_k - \mu^t_l) e_{kl} =: \mu^t_{kl} e_{kl}, \quad (e^a_{kl})^* = e^a_{lk}.
	\end{equation}
	The crucial observation is that these eigen\emph{vectors} are independent of $t$.
	Now,
	\begin{equation} \label{h_f_w}
		H_F = -\gamma^t_{\alpha\beta} M^t_{ab} \theta_{\alpha a} \theta_{\beta b} 
			= W_{\alpha a, \beta b}\ \theta_{\alpha a} \theta_{\beta b},
	\end{equation}
	where $W := -\sum_t \gamma^t \otimes M^t$. From the above observations we have the
	ansatz $E_{\mu kl} := v_{\mu} \otimes e_{kl}$ for the eigenvectors of $W$, giving
	\begin{equation} \label{w_eigenvectors}
		\textstyle
		W E_{\mu kl} = -\sum_t \gamma^t v \otimes M^t e_{kl} = \gamma(k,l) v_{\mu} \otimes e_{kl},
	\end{equation}
	where $\gamma(k,l) := -\sum_t \mu^t_{kl} \gamma^t$ squares to $\sum_t (\mu^t_{kl})^2$.
	Letting $v_{\mu} = v_{\pm jkl}$ denote the corresponding 16 eigenvectors of $\gamma(k,l)$, we find
	\begin{equation} \label{w_eigenvalues}
		\textstyle
		W E_{\pm jkl} = \pm \sqrt{\sum_t (\mu^t_{kl})^2} E_{\pm jkl}
	\end{equation}
	and $H_F$ therefore has $E_0 := -16 \sum_{k<l} \sqrt{\sum_{t=1}^9 (\mu^t_k - \mu^t_l)^2}$ as its lowest eigenvalue.
	
	This agrees with the following two previously known cases: 
	\cite{hasler_hoppe}, where
	only $X_9$ is assumed to have large eigenvalues so that 
	$E_0 \to -16 \sum_{k<l} |\mu^9_k - \mu^9_l|$; 
	as well as the SU(2)-case studied above, where
	\eqref{coords} with e.g. $e_A = \delta_{A3}$
	gives $E_0 = -16r$.

\section*{Acknowledgements}

	One of us (J.H.) would like to 
	thank Choonkyu Lee for his hospitality,
	and Ki-Myeong Lee for useful discussions.

\noindent
V. Bach, FB Mathematik, Universit\"at Mainz, DE-55099 Mainz, Germany\\
\phantom{ }\\
\noindent
J. Hoppe and D. Lundholm, Department of Mathematics, Royal Institute of Technology, SE-10044 Stockholm, Sweden


\begin{thebibliography}{99}
	\bibitem{hoppe}
	J. Hoppe, \textit{On the construction of zero energy states in supersymmetric matrix models I, II, III,} hep-th/9709132, 9709217, 9711033.
	\bibitem{halpern_schwartz}
	M.B. Halpern, C. Schwartz, \textit{Asymptotic search for ground states of SU(2) matrix theory.} Int. J. Mod. Phys. A 13 (1998) 4367, hep-th/9712133.
	\bibitem{graf_hoppe}
	G.M. Graf, J. Hoppe, \textit{Asymptotic ground state for 10-dimensional reduced supersymmetric SU(2) Yang-Mills theory,} hep-th/9805080.
	\bibitem{froehlich_et_al}
	J. Fr\"ohlich, G.M. Graf, D. Hasler, J. Hoppe, S.-T. Yau, \textit{Asymptotic form of zero energy wave functions in supersymmetric matrix models,} Nucl. Phys B 567 (2000), 231-248.
	\bibitem{hasler_hoppe}
	D. Hasler, J. Hoppe, \textit{Asymptotic factorisation of the ground-state for SU($N$)-invariant supersymmetric matrix-models,} hep-th/0206043.
\end{thebibliography}
\end{document}